\newif\iftwocolumns
\twocolumnstrue       % JOURNAL STYLE

\iftwocolumns
  \documentclass[aps,prl,twocolumn,groupedaddress,amsmath,amssymb]{revtex4}
\else
  \documentclass[aps,prl,preprint,groupedaddress,amsmath,amssymb]{revtex4}
\fi
\usepackage{graphicx}% Include figure files
\usepackage{dcolumn}% Align table columns on decimal point
\begin{document}

%\preprint{}


\title{Comment on `Statistical mechanics of developable ribbons' by
L.~Giomi and L.~Mahadevan}


\author{E.~L.~Starostin}
%\email{e.starostin@ucl.ac.uk}
%\email{eugene.starostin@daad-alumni.de}
%\thanks{}
\author{G.~H.~M.~van~der~Heijden}
%\email{g.heijden@ucl.ac.uk}
%used with the \author command). \
\affiliation{Centre for Nonlinear Dynamics, University College London,
Gower Street, London WC1E 6BT, UK}

\date{\today}

%\pacs{}
%\keywords{}

\maketitle

%\section{}
%\subsection{}
%\subsubsection{}

\noindent
Giomi and Mahadevan \cite{Giomi10}
claim to provide evidence for ``an underlying helical
structure" of ribbon-like polymers even in the absence of intrinsic
curvature (or, in the authors' words, ``of a preferential zero-temperature
twist"). They find a persistence length that is over 3 times that of a
wormlike chain having the same bending rigidity. We show that these
results are an artefact of the authors' flawed formulation, which produces
a helical bias, and that no support for the described effect can be claimed.

The bias stems from the authors' use of the Frenet frame (FF) in place of a
continuous material frame (MF). FF is known to be an improper choice for
ribbons~\cite{Rappaport07}. Although FF
seems natural for developable strips, as the principal normal to the
centreline coincides with the normal to the surface, it has the property
that it flips at inflection points of the centreline, i.e., the principal
normal and binormal change sign, becoming opposite to the continuous
vectors of MF (with $\phi$ jumping $\phi\to\phi+\pi$).
The centreline of the ribbon is modelled as a polygonal line with
neighbouring vertices ${\bf x}_i$ at distance $a$.
The shape of the ribbon between ${\bf x}_{i-1}$ and ${\bf x}_{i}$
may be inflected or not. If it is, then the MF, if aligned with the FF
at ${\bf x}_{i-1}$, will be opposite to the FF at ${\bf x}_{i}$
and its discrete torsion is
$\tau^2=a^{-2}|{\bf b}_i+{\bf b}_{i-1}|^2=2a^{-2}(1+\cos\phi_i)$.
Consequently, for the inflected shape the elastic energy is smaller if
$\phi_i>\frac{\pi}{2}$. However, this option is ignored in~\cite{Giomi10}
where MF is always coincident with FF at ${\bf x}_i$, so it is
always assumed that the ribbon segment has no inflection. Instead, it
must be highly bent with a high localised torsion to match the boundary
conditions at ${\bf x}_{i-1}$ and ${\bf x}_{i}$. It means that in the
continuum limit $a\to 0$ the torsion diverges. The employed discretisation
prevents inflected solutions to be approximated.

Inflected conformations are therefore excluded in the Monte Carlo
simulations, which generate an ensemble of ribbon shapes with an
artificial bias in favour of solutions without inflection points.
One should expect the absence of inflected conformations (see Fig.~4) to
affect the ensemble's statistical properties: experimenting with a strip
of paper it is easy to produce inflected shapes; they don't cost much
energy. Consider two deformations of the strip: a strip locally held in a
shape close to a ring and a strip held in an approximate S-shaped form.
The centreline of the former has no inflection point, contrary to the
latter.
Clearly the elastic energy is about the same in these two configurations,
yet the procedure of~\cite{Giomi10} completely ignores the inflected
S-shape.

The bias can be eliminated by using a MF throughout. For instance,
sticking with the present angles, rather than through a jump in $\phi$
one can describe inflection points by allowing $\theta$ to take
on negative values. The range of the angles would then change to
$\theta \in (-\pi, \pi], \phi \in (-\frac{\pi}{2}, \frac{\pi}{2}]$,
also used for discrete chains in \cite{Hu11}. This eliminates a possible
jump by $\pi$ in $\phi$ and hence amounts to using a non-flipping, material,
frame. Following the authors' reasoning one would then assume that
$\langle\sin\theta\rangle=0=\langle\sin\phi\rangle$, leading to three real
eigenvalues of the averaged transfer matrix $\boldsymbol{R}$ and thus
exponential decay of the tangent-tangent correlation function
$\langle\boldsymbol{t}_n\cdot\boldsymbol{t}_0\rangle$, in contradiction
to the oscillatory decay shown in Fig.~3.

The authors also present an analytical calculation that confirms their
numerical results by producing a persistence length $l_p$ for developable
ribbons that is more than 3 times that of a wormlike chain of the same
bending rigidity. However, for the redefined angles, with area element
$\mbox{d} \Omega = |\sin\theta| \mbox{d}\theta \mbox{d}\phi$, following
the computations in \cite{Giomi10} we arrive at a modified partition
function, Eq.~(9), in which $I_0(\beta\mu)$ is replaced by
$I_0(\beta\mu)+L_0(\beta\mu)$, where $L_0$ is the modified Struve function
of 0th order. For $\zeta \gg 1$,
$L_0(\zeta)=I_0(\zeta)-\frac{2}{\pi\zeta}+{\cal O}(\zeta^{-3})$,
which implies that the asymptotic expression for $\langle\cos\theta\rangle$ 
on p.~4 (left, middle) of~\cite{Giomi10} remains valid. Therefore, we find
$l_p=-a(\log\langle\cos\theta\rangle)^{-1} \simeq \frac{32}{35} \frac{Dw}{k_B T}$,
which is slightly lower than for the wormlike chain model.

We conclude that there is no ground for the claim in \cite{Giomi10} that
developable ribbons possess an ``underlying helical structure". This
conclusion is also in agreement with calculations in \cite{Mergell02}, where
a properly discretised developable ribbon is found to yield an exponentially
decaying tangent-tangent correlation function, consistent with our result
above.

\end{document}